\documentstyle[aps,preprint]{revtex}

\begin{document}

\input epsf
\newcommand{\infig}[2]{\begin{center}\mbox{ \epsfxsize #1
                       \epsfbox{#2}}\end{center}}

\newcommand{\be}{\begin{equation}}
\newcommand{\nn}{\nonumber}
\newcommand{\ee}{\end{equation}}
\newcommand{\bea}{\begin{eqnarray}}
\newcommand{\eea}{\end{eqnarray}}
\newcommand{\wee}[2]{\mbox{$\frac{#1}{#2}$}}   
\newcommand{\unit}[1]{\,\mbox{#1}}
\newcommand{\degree}{\mbox{$^{\circ}$}}
\newcommand{\ltish}{\raisebox{-0.4ex}{$\,\stackrel{<}{\scriptstyle\sim}$}}
\newcommand{\vs}{{\em vs\/}}
\newcommand{\bin}[2]{\left(\begin{array}{c} #1 \\ #2\end{array}\right)}
\newcommand{\pred}{^{\mbox{\small{pred}}}}
\newcommand{\retr}{^{\mbox{\small{retr}}}}
\newcommand{\p}{_{\mbox{\small{p}}}}
\newcommand{\m}{_{\mbox{\small{m}}}}
\newcommand{\rs}[1]{_{\mbox{\small{#1}}}}	
\draft

\title{Retrodiction with two-level atoms: atomic previvals}

\author{John Jeffers${}^1$, Stephen M. Barnett${}^1$ and David T. Pegg${}^2$}
\address{${}^1$ Department of Physics and Applied Physics, University
of Strathclyde, Glasgow G4 ONG, United Kingdom.\\ ${}^2$ Faculty of
Science, Griffith University, Nathan, Brisbane, Q111, Australia.}

\date{\today}
\maketitle

\begin{abstract}
In the Jaynes-Cummings model a two-level atom interacts with a
single-mode electromagnetic field. Quantum mechanics predicts collapses
and revivals in the probability that a measurement will show the atom
to be excited at various times after the initial preparation of the
atom and field.  In retrodictive quantum mechanics we seek the
probability that the atom was prepared in a particular state given the
initial state of the field and the outcome of a later measurement on
the atom.  Although this is not simply the time reverse of the usual
predictive problem, we demonstrate  in this paper that retrodictive
collapses and revivals also exist.  We highlight the differences
between predictive and retrodictive evolutions and describe an
interesting situation where the prepared state is essentially
unretrodictable.
\end{abstract}


\narrowtext


\section{Introduction}

The rapid development of quantum information theory in recent years has
given fresh impetus to the study of retrodictive quantum theory. A
retrodictive quantum formalism was first proposed more than 30 years
ago, and others have followed \cite{oldret}. Recently its utility has
been extended by the application of Bayes' theorem \cite{bayes} to the
conditional probabilities derived using predictive quantum theory
\cite{newret}. The theory applies to closed systems, and also to open
systems in which the system of interest interacts with an unmeasured
environment \cite{atomic,ampatt,master}.

Normally we want to predict the future based on our knowledge of the
present, so in predictive quantum theory the state of the system at any
time between preparation and measurement is the evolved prepared
state.  Sometimes, however, our knowledge of initial states is not
complete. If we know the result of a measurement of the state we can
assign a retrodictive state on the basis of the measurement outcome.
Each measurement outcome has associated with it a probability operator
measure (POM) element\cite{pom}. It is possible to prove using Bayes'
theorem \cite{bayes} that the retrodictive density operator is simply
the normalised POM element \cite{newret}. For closed systems this state
evolves backwards in time to the preparation time according to the
Schr\"odinger equation, when it collapses on to one of a set of
possible initially prepared states. Normally the predictive and
retrodictive states assigned to a system at a particular time between
preparation and measurement will be different.

For open systems the evolution is more complicated. The simple
time-reversal property of closed systems does not apply.  In general
the system to be measured interacts with another unmeasured system
usually called the environment, which is traced out of the problem to
give the (nonunitary) evolution of the system of interest. This
evolution is governed by a master equation for the system density
operator. We have recently derived methods for solving retrodictive
problems in open systems based on standard predictive master equation
techniques \cite{atomic}. Furthermore, we have derived a retrodictive
master equation which traces the evolution of such systems backwards in
time \cite{master}.

These master equation methods have been used to prove that optical
amplifiers and attenuators are predictive/retrodictive inverses of one
another \cite{ampatt}. Also, retrodiction from measured atomic states,
for a two-level atom interacting with an environment which consists of
a multimode electromagnetic field, has been studied \cite{atomic}. This
system illustrates various general principles of retrodictive open
systems.  In particular, if nothing is known about the initial state of
the system the retrodictive steady state is usually the no-information
state. This is an equally-weighted mixture of all the possible input
states. In addition, the retrodictive decay rate depends upon the
measurement outcome. There are also properties peculiar to this
two-level system.  If the atom is illuminated with coherent light
retrodictive Rabi oscillations occur, which decay eventually to the
no-information state.

In this paper we concentrate in particular on the two-level atom
interacting with a single cavity mode of the quantised electromagnetic
field. In standard predictive quantum theory, if the evolution of the
atomic state is driven by a known coherent field, Rabi oscillations
occur both in the atomic population, and in the off-diagonal atomic
coherences. The frequency of the oscillations increases with the square
root of the number of quanta of energy in the system. For a coherent
driving field the number of photons is not completely certain; rather
the field is in a weighted superposition of all photon number states.
After a short time the different Rabi oscillations for each of the
number states within the superposition get out of phase with one
another and the oscillations of the whole system collapse.  Later, when
sufficient time has elapsed so that the oscillations get back into
phase, there is a revival of the oscillating atomic population
\cite{eberly,Gea,shore}. Here we look at this system from the
retrodictive point of view, where we measure the final state of the
atom knowing nothing about its initial state. The field is initially in
a coherent state, but it is not measured after the interaction. The
retrodictive situation is not the time-reverse of the predictive
situation because we have different knowledge in the two cases. In the
retrodictive situation we know the initial state of the field and the
final measured state of the atom, whereas in the predictive case these
states are both known at the initial time. This poses an interesting
question. Would we retrodict that the retrodictive Rabi oscillations
were in phase at times prior to the measurement of the field state,
that is, would we retrodict any ``previvals"?

The paper is organised as follows. In Section (II) we briefly describe
the general features of retrodictive quantum theory, and its
application to the two-level atom interacting with an electromagnetic
field. We apply this in Section (III) and give results for the
retrodictive density matrix and preparation probabilities. Section (IV)
contains a summary and discussion of the main results of the paper.


\section{Prediction and retrodiction}
\label{sec:predret}

In this section we provide brief details of retrodictive quantum
theory. A fuller account can be found in references \cite{newret} and
\cite{atomic}. Suppose that we have a preparation device which produces
output states $\hat{\rho}\pred_i$ with prior probabilities $P(i)$,
where $\hat{\rho}\pred_i$ is the usual density operator of predictive
quantum mechanics. This state can evolve and interact with other
systems until it is measured by a measuring device. A general
description of a measurement is given by a measurement POM \cite{pom}.
This is a set of non-negative definite, Hermitian elements
$\hat{\Pi}_j$ which sum to the unit operator, each element
corresponding to a particular measurement outcome. In general there is
no requirement that there be the same number of POM elements as there
are states which span the system space, but for von Neumann measurement
this is so, and the POM elements are simply the projectors of the
particular chosen states which span the space.  Suppose that
preparation takes place at time $t\p$ and measurement at a later time
$t\m$. Within this framework the predictive probability that the
measurement outcome $\hat{\Pi}_j$ is obtained given that the state
$\hat{\rho}\pred_i$ was prepared is
\bea
\label{predconprob}
P(j|i)=\mbox{Tr}\left(\hat{\rho}\pred_i (t\m) \hat{\Pi}_j\right),
\eea
where 
\bea
\label{rhopredclosed}
\hat{\rho}\pred_i(t\m)= \hat{U}(\tau)
\hat{\rho}\pred_i(t\p) \hat{U}^\dagger(\tau)
\eea is the evolved initial density operator and 
\bea
\label{evolve}
\hat{U}(\tau) = \exp \left(-\frac{i}{\hbar} \hat{H} \tau\right)
\eea
is the evolution operator, which operates for the length of time
between preparation and measurement, $\tau = t\m-t\p$.

Suppose that instead of calculating the predictive probability $P(j|i)$
we wish to calculate the retrodictive conditional probability $P(i|j)$
that the state $\hat{\rho}\pred_i$ was prepared, given our measurement
result $\hat{\Pi}_j$. It is possible to do this by calculating all
possible predictive conditional probabilities for the system, and then
using Bayes' theorem. A simpler and more natural approach is to use
retrodictive quantum theory, so that the required probability can be
written \cite{newret,atomic}
\bea
\label{retconprob}
P(i|j)= \frac {\mbox{Tr} \left[ \hat{\Lambda}_i \hat{\rho}\retr_j
(t\p) \right]} {\mbox{Tr} \left[ \hat{\Lambda}
\hat{\rho}\retr_j (t\p) \right]}.
\eea Here the operator $\hat{\Lambda}_i$ is the preparation device
operator, and
\bea
\hat{\Lambda}=\sum_i \hat{\Lambda}_i = \sum_i P(i)
\hat{\rho}\pred_i,
\eea is the {\it a priori} density operator, the sum of each possible
preparation density operator weighted by its prior probability of
production.  $\hat{\Lambda}$ is the best description of the state we
can give without knowing the outcome of the preparation or
measurement.  The retrodictive density operator at the preparation time
is simply the normalised measurement POM element evolved back from the
measurement time to the preparation time,
\bea
\label{rhoretrclosed}
\hat{\rho}\retr_j (t\p) = \hat{U}^\dagger(\tau)
\hat{\rho}\retr_j(t\m) \hat{U}(\tau),
\eea with
\bea
\hat{\rho}\retr_j(t\m) =
\frac{\hat{\Pi}_j}{\mbox{Tr}\hat{\Pi}_j}.
\eea

The above formulae for the conditional probabilities (eqs.
(\ref{predconprob}) and (\ref{retconprob})) apply equally well for open
systems, where the system of interest interacts with an unmeasured
environment with many degrees of freedom. If this environment causes
information to be lost about the system, and the Born-Markov
approximation holds, then the the evolved density operators are the
solutions of master equations \cite{jch}. In eq.(\ref{predconprob}) the
density operator required for such an open system, which for a closed
system would be given by eq. (\ref{rhopredclosed}), is the solution of
the usual master equation forwards in time from the preparation  time
to the measurement time.  However, in eq.(\ref{retconprob}) the
solution required instead of eq.  (\ref{rhoretrclosed}) is that of the
retrodictive master equation, giving the evolution backwards in time
from the measurement time to the preparation time. We have recently
derived this equation from the general principle that conditional
probabilities should be independent of the time of collapse of the
wavefunction\cite{master}.  In the system considered in the present
paper, however, the Born-Markov approximation {\it is not made}, and so
we must consider the full evolution of the coupled atom-field system.
\section{Retrodiction for the coupled atom-field system}

Here we apply the retrodictive formalism to a coupled system
consisting of a two-level atom and a single cavity mode of an
electromagnetic field. The interaction between an atom with upper level
$|e\rangle$ and lower level $|g \rangle$, and an electromagnetic field
is governed by the Jaynes-Cummings Hamiltonian \cite{jch}. In the
interaction picture this is
\bea
\label{jcham}
\hat{H}= \frac{\hbar \Delta}{2} \hat{\sigma}_3-i \hbar \lambda
\left(\hat{\sigma}_+\hat{a}-\hat{a}^\dagger \hat{\sigma}_- \right),
\eea 
where $\Delta$ is the detuning between the atomic frequency and the
light, $\hat{\sigma}_3=|e\rangle \langle e|-|g\rangle \langle g|$ is
the atomic inversion operator, $\hat{\sigma}_+ = |e\rangle \langle g|$
and $\hat{\sigma}_-= |g\rangle \langle e|$ are the atomic raising and
lowering operators, $\hat{a}^\dagger$ and $\hat{a}$ are the creation
and annihilation operators for the single mode field, and $\lambda$ is
the coupling constant. The rotating wave approximation, which has been
made in deriving this Hamiltonian, ensures that whenever a photon is
lost from the field the atomic state must change from $|g\rangle$ to
$|e\rangle$. In the standard predictive picture of quantum mechanics a
coupled atom-field system evolves forwards in time according to this
Hamiltonian from a preparation time $t\rs{p}$ to a measurement time
$t\rs{m}$.  After this has happened the coupled density operator for
the whole system is
\bea
\label{coupred}
\hat{\rho}\pred \rs{af}(t\rs{m}) = \hat{U}(\tau) \hat{\rho}\pred
\rs{a}(t\p) \otimes \hat{\rho}\pred \rs{f}(t\p) \hat{U}^\dagger(\tau),
\eea 
where $\hat{\rho}\pred \rs{af}(t\m)$ is the coupled density operator
for the atom-field system at the measurement time, $\hat{\rho}\pred
\rs{a}(t\p)$ and $\hat{\rho}\pred \rs{f}(t\p)$ are the uncoupled atom
and field density operators at the preparation time. $\hat{U}(\tau)$ is
the evolution operator for the coupled system, given by
eq.(\ref{evolve}) with Hamiltonian given by eq. (\ref{jcham}).  If the
field is unmeasured the atomic state is simply found by tracing over
field states and vice versa,
\bea
\label{rhoa}
\hat{\rho}\pred \rs{a}(t\m) &=& \mbox{Tr}\rs{f}\left[\hat{U}(\tau)
\hat{\rho}\pred \rs{a}(t\p) \otimes \hat{\rho}\pred \rs{f}(t\p)
\hat{U}^\dagger(\tau)\right],\\
\label{rhof}
\hat{\rho}\pred\rs{f}(t\m) &=& \mbox{Tr}\rs{a}\left[\hat{U}(\tau)
\hat{\rho}\pred \rs{a}(t\p) \otimes \hat{\rho}\pred \rs{f}(t\p)
\hat{U}^\dagger(\tau)\right].
\eea 
Alternatively, we can condition the field by measuring the atom, or the
atom by measuring the field. This is done using the POM element
corresponding to the measurement outcome. Thus the conditioned atomic
and field states immediately after the measurement are
\bea
\label{rhoapred}
\hat{\rho}\pred\rs{a}(t\m) &\propto&
\mbox{Tr}\rs{f}\left[\hat{\Pi}\rs{f}\hat{U}(\tau) \hat{\rho}\rs{a}(t\p)
\otimes \hat{\rho}\rs{f}(t\p) \hat{U}^\dagger(\tau)\right],\\
\label{rhofpred}
\hat{\rho}\pred\rs{f}(t\m) &\propto&
\mbox{Tr}\rs{a}\left[\hat{\Pi}\rs{a}\hat{U}(\tau) \hat{\rho}\rs{a}(t\p)
\otimes \hat{\rho}\rs{f}(t\p) \hat{U}^\dagger(\tau)\right],
\eea
where $\hat{\Pi}\rs{a(f)}$ is the POM element corresponding to the
outcome of the measurement performed on the atom (field).

The retrodictive picture differs from above in that the state of
the system is assigned on the basis of the measurement outcome. Thus if
the measurement POM elements for the atom and the field are
$\hat{\Pi}\rs{a}(t\m)$ and $\hat{\Pi}\rs{f}(t\m)$, the coupled initial
density operator corresponding to equation (\ref{coupred}) is
\bea
\label{coupret}
\nonumber \hat{\rho}\retr \rs{af}(t\p) &=& \hat{U}^\dagger(\tau)
\hat{\rho}\retr \rs{a}(t\m) \otimes \hat{\rho}\retr \rs{f}(t\m)
\hat{U}(\tau)\\
&\propto& \hat{U}^\dagger(\tau) \hat{\Pi}\rs{a}(t\m)
\otimes \hat{\Pi}\rs{f}(t\m) \hat{U}(\tau).
\eea 
The atom or the field will in general have been prepared in one of a
set of initial states, and this state conditions the coupled density
operator similarly to equations (\ref{rhoapred}) and (\ref{rhofpred}),
to give retrodictive density operators for the atom and field
respectively
\bea
\label{rhoaretr}
\hat{\rho}\retr \rs{a}(t\p) &\propto&
\mbox{Tr}\rs{f}\left[\hat{\rho} \pred \rs{f}(t\p)\hat{U}^\dagger(\tau)
\hat{\Pi} \rs{a}(t\m) \otimes \hat{\Pi}\rs{f}(t\m)
\hat{U}(\tau)\right],\\
\label{rhofretr}
\hat{\rho}\retr \rs{f}(t\p) &\propto& \mbox{Tr}\rs{a} \left[\hat{\rho}
\pred \rs{a}(t\p) \hat{U}^\dagger(\tau) \hat{\Pi}\rs{a}(t\m) \otimes
\hat{\Pi}\rs{f}(t\m) \hat{U}(\tau)\right],
\eea
where the constant of proportionality is determined by normalisation.
If there is no information at all about the preparation of the initial
states then $\hat{\rho}\pred \rs{a}(t\p)$ and $\hat{\rho}\pred
\rs{f}(t\p)$ become proportional to the unit operators for the atom and
the field.


\section{Retrodiction of the atomic state: collapses and previvals}
\label{sec:previvals}

Here we consider a particular situation in which retrodiction is a
useful tool for finding atomic states at an earlier time. We assume
that the initial atomic state is completely unknown, but the initial
field is in a known coherent state. A measurement is made of the atomic
state, and this result is used as a basis for retrodicting the atomic
state at the preparation time. The retrodictive atomic density operator
is given by equation (\ref{rhoaretr}). For ease of calculation we
assume that the atomic state is measured to be in one of a pair of
orthogonal states which span the two-level atomic basis, for example,
the excited or ground state. We can calculate quantities based on other
assumed bases using results for this one. The associated atomic POM
element has unit trace and so the retrodictive atomic density operator
immediately prior to measurement is simply the measurement POM element
itself. Initially the field is in a coherent state, but the final state
is unmeasured so the field measurement POM element is simply the unit
operator for the field \cite{unitfield}. Thus equation (\ref{rhoaretr})
gives the initial retrodictive atomic density operator as
\bea
\label{atomstate}
\hat{\rho}\retr \rs{a}(t\p) \propto \langle \alpha |
\hat{U}^\dagger(\tau) \hat{\Pi}\rs{a}(t\m) \otimes \hat{1}\rs{f}
\hat{U}(\tau) |\alpha \rangle,
\eea
where $|\alpha \rangle$ is the coherent field state and $\hat{1}\rs{f}$
is the unit state. It is relatively straightforward to compute the
density matrix elements,
\bea
\label{atomelement}
\langle l| \hat{\rho}\retr _{\mbox{\small{a}}}(t\p) |m \rangle \propto
\langle \alpha | \langle l| \langle  \hat{U}^\dagger(\tau)
\hat{\Pi}\rs{a}(t\m) \otimes \hat{1}\rs{f} \hat{U}(\tau) |m \rangle
|\alpha \rangle ,
\eea
where $l$ or $m$ can be either the ground or excited states, or one of
any other pair of orthogonal states which span the two-level atomic
basis. In fact $\hat{U}(\tau) |m \rangle |\alpha \rangle $ is simply the
state that the atomic state $|m\rangle$ would have evolved into after a
time $\tau$ in the predictive formalism. This allows us to use the
well-known solution to the Jaynes-Cummings model \cite{jch},
\bea
\hat{U}(\tau) |j \rangle |\alpha \rangle = \sum_{n=0}^\infty \left[
c_{g,n}(\tau) | g \rangle | n \rangle + c_{e,n}(\tau) | e \rangle | n
\rangle \right],
\eea
where the states $|n \rangle$ are the photon number states and the
ground and excited state amplitudes depend upon the coherent state
expansion coefficients in the number state basis. The amplitudes are
found to be
\bea
\nonumber c_{g,n}(\tau) &=& c_{g,n}(0)\left[ \cos{\frac{\Omega(n)
\tau}{2} } + \frac{i\Delta}{\Omega(n)} \sin{\frac{\Omega(n) \tau}{2}}
\right] \\ &+& c_{e,n-1}(0) \frac{2\lambda n^{1/2}}{\Omega(n)}
\sin{\frac{\Omega(n) \tau}{2}},\\
\nonumber c_{e,n-1}(\tau) &=& c_{e,n-1}(0)\left[ \cos{\frac{\Omega(n)
\tau}{2}} - \frac{i\Delta}{\Omega(n)} \sin{\frac{\Omega(n) \tau}{2}}
\right] \\ &-& c_{g,n}(0) \frac{2\lambda n^{1/2}}{\Omega(n)}
\sin{\frac{\Omega(n) \tau}{2}},
\eea
where $\Omega(n) = (\Delta^2 +4\lambda^2 n)^{1/2}$ is the Rabi
frequency, and $c_{g,n}(0)$ and $c_{e,n}(0)$ are the initial
amplitudes, given by
\bea
c_{g,n}(0)&=&a_n c_g(0)\\
c_{e,n}(0)&=&a_n c_e(0).
\eea
These initial values are proportional to the number state expansion
coefficients of the coherent state,
\bea
a_n=\exp{(-|\alpha|^2/2)}\frac{\alpha^n}{\sqrt{n}}.
\eea We can use these formulae to calculate retrodictive matrix
elements and probabilities, given that the atom was measured to be in a
particular state. The calculations are relatively straightforward, and
details are omitted.

We consider  the case where the atom is known to have been prepared
either in the excited state or the ground state with equal {\it a
priori} probabilities. The appropriate preparation device operators in
(\ref{retconprob}) are $|e\rangle \langle e|/2$ and $|g\rangle \langle
g|/2$. Figure 1 shows a typical plot of the retrodictive conditional
probability, in this case the probability that the atom was prepared in
the ground state given that it was measured in the excited state.
It shows retrodictive Rabi oscillations
which collapse as the elapsed time before the measurement increases. If
the preparation time was long enough before the measurement time the
oscillations are seen to revive, just as in the predictive case. The
characteristic revival time is the same as the predictive one. Thus
``previvals" or ``earlier revivals" in the preparation probability do
exist.

Given that both the collapse time and the revival time are identical
for the predictive and retrodictive evolutions one might think that the
retrodictive evolution is simply the time-reverse of the predictive
evolution. This is not the case, as is shown in figure 2, which depicts
the evolution for a weaker coherent state. Here we compare predictive
and retrodictive evolutions. Figure 2(a)
shows the retrodictive conditional probability that the ground state
was prepared given that the atom has been measured in the excited
state. This can be compared with figures 2(b) and 2(c). These show,
respectively, the predictive conditional probabilities that (b) the
atom is measured in its ground state given that the excited state was
prepared, and (c) the atom is measured in its excited state given that
the ground state was prepared.  The retrodictive evolution is clearly
the time-reverse of neither of these two predictive evolutions .

This point can be illustrated more dramatically by considering other
measured states. For example, for a high-amplitude coherent state,
after (predictive) Rabi oscillations collapse, at a time
$\tau=\pi/(2\Omega(\overline{n}))$  the state of the system
approximately factorises into uncoupled atomic and field states
\cite{Gea}. No matter what the initial prepared atomic state, after
this period, which is half the revival time, the atom is prepared by
the system in the state
\bea
\label{measup}
|-\rangle = \frac{1}{\sqrt{2}}\left( |g\rangle - e^{i\phi} |e\rangle
\right),
\eea 
where $\phi$ is the phase of the coherent state amplitude $\alpha$
\cite{state}. The predictive evolution of the state then consists of
the revival of the Rabi oscillations.

On the other hand, if we measure the atom in this superposition state
and try to retrodict the prepared state, the evolution is completely
different. This is illustrated by figure 3, which is a plot of the
excited state preparation probability. For
a short delay there are Rabi oscillations which collapse, leaving a
slow oscillation whose period is associated with the revival time of
the system. If the delay is equal to half the revival time then the
excited state probability passes through the value $1/2$. The atomic
density operator becomes
\bea
\label{noinf}
\hat{\rho} \retr \left( \frac{\pi}{2\Omega(\overline{n})} \right) =
\frac{1}{2}\left[ |e\rangle \langle e| + |g\rangle \langle g| \right],
\eea
the no-information state. We say that the atomic state is {\it
unretrodictable} at this time. The reason for this is simply that our
measurement of the atom only provides information about the field at
this time.



\section{Conclusions}
\label{sec:concs}
In this paper we have analysed the two-level atom interacting with a
single-mode electromagnetic field in a coherent state from a
retrodictive point of view. This system shows predictive collapses and
revivals in the atomic state probabilities \cite{eberly}. We have
demonstrated the existence of retrodictive collapses and previvals of
the Rabi oscillations in the atomic state probabilities. This follows
on from our previous work which demonstrated the existence of
retrodictive Rabi oscillations \cite{atomic}.
 
We have shown that the retrodictive and predictive evolutions are
different. The differences are most marked when either of two
particular criteria are satisfied.  Firstly, when the mean number of
photons is low, for all measured atomic states, it becomes easy to
differentiate between the predictive and retrodictive evolutions.
Secondly, for high mean photon number, if the measured state is that to
which all prepared states decay after a particular time then the
retrodictive evolution takes on a strange character. There is a
low-frequency retrodictive oscillation in the atomic state probability
with a period equal to twice the revival time.  Furthermore, when the
time between preparation and measurement is equal to half the revival
time the retrodictive state is unretrodictable. The probability that
any one of a pair of states which span the atomic space was prepared
will then be one half.


\acknowledgments

The authors would like to thank the United Kingdom Engineering and
Physical Sciences Research Council and the Australian Research Council
for financial support. SMB thanks the Royal Society of Edinburgh and
the Scottish Office Executive Education and Lifelong Learning
Department for the award of a Support Research Fellowship.


\newpage
\section*{Figure Captions}

Figure 1:  A plot of the retrodictive conditional probability that the
atom was prepared in the ground state given a later measurement in the
excited state as a function of the normalised difference between
preparation and measurement times $\lambda \tau$. Parameters: detuning
$\Delta=0$ and coherent state amplitude $\alpha=5$.

Figure 2: (a) Same plot as figure 1, but with coherent state amplitude
$\alpha=1.4$.
(b) A plot of the predictive conditional probability that the atom was
measured in the ground state given that it was prepared in the excited
state as a function of $\lambda \tau$. Parameters are as for (a).
(c) A plot of the predictive conditional probability that the atom was
measured in the excited state given that it was prepared in the ground
state as a function of $\lambda \tau$. Parameters are as for (a).

Figure 3: A plot of the retrodictive conditional probability that the
atom was prepared in the excited state given a later measurement in the
superposition state $\frac{1}{\sqrt{2}}\left( |g\rangle - e^{i\phi}
|e\rangle \right)$ as a function of $\tau$. The parameters are as
in figure 1.


\begin{thebibliography}{99}
\bibitem{unitfield} This is because the act of not measuring the field
is equivalent to performing a measurement with only one possible
outcome. The probability for this outcome is, of course, unity for all
possible states and so the POM element is the identity
\bibitem{state} This state is not the one found in reference
\cite{Gea}, a difference which can be explained by the fact that our
Hamiltonian has a slightly different form.
\end{thebibliography}
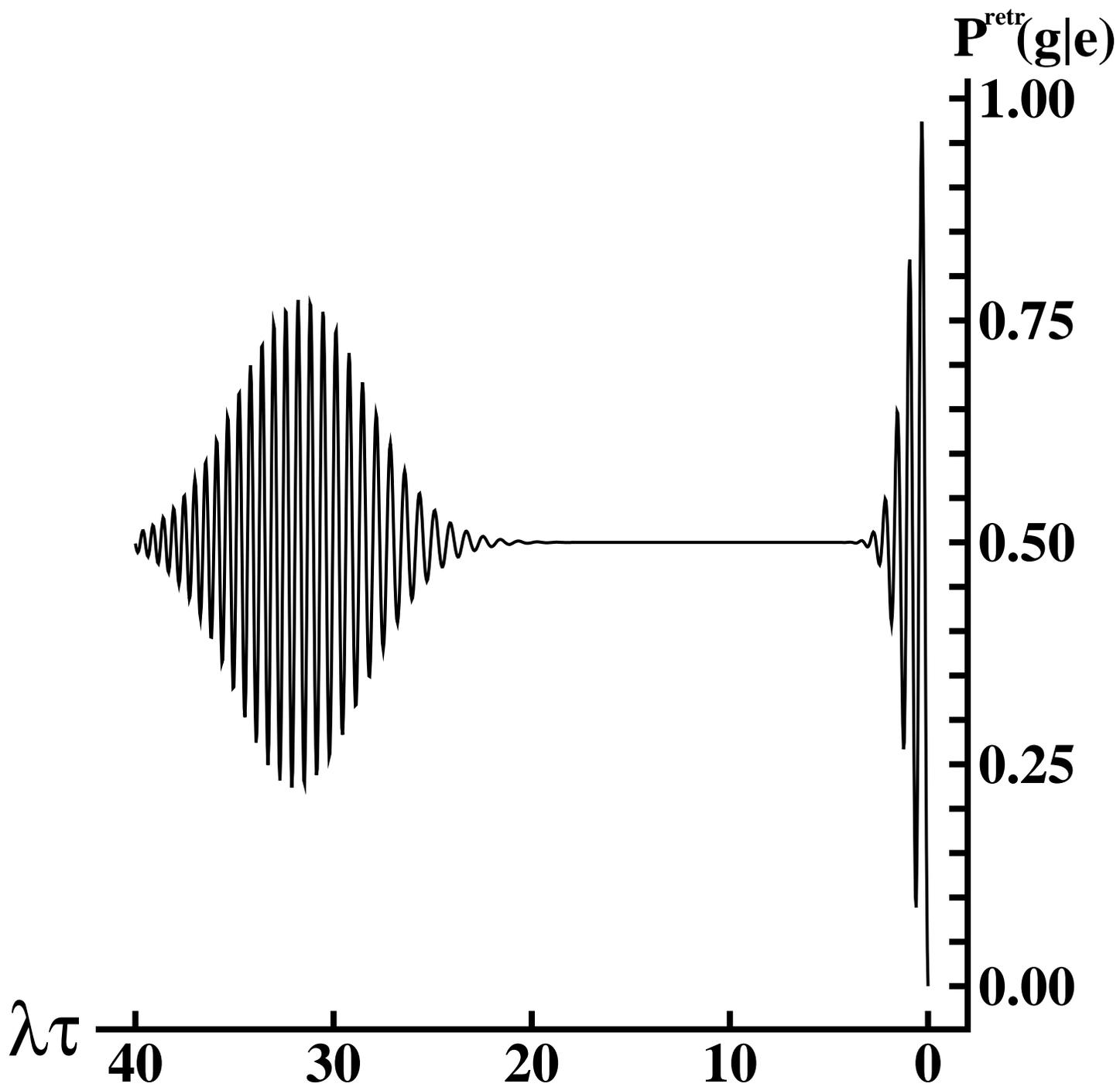
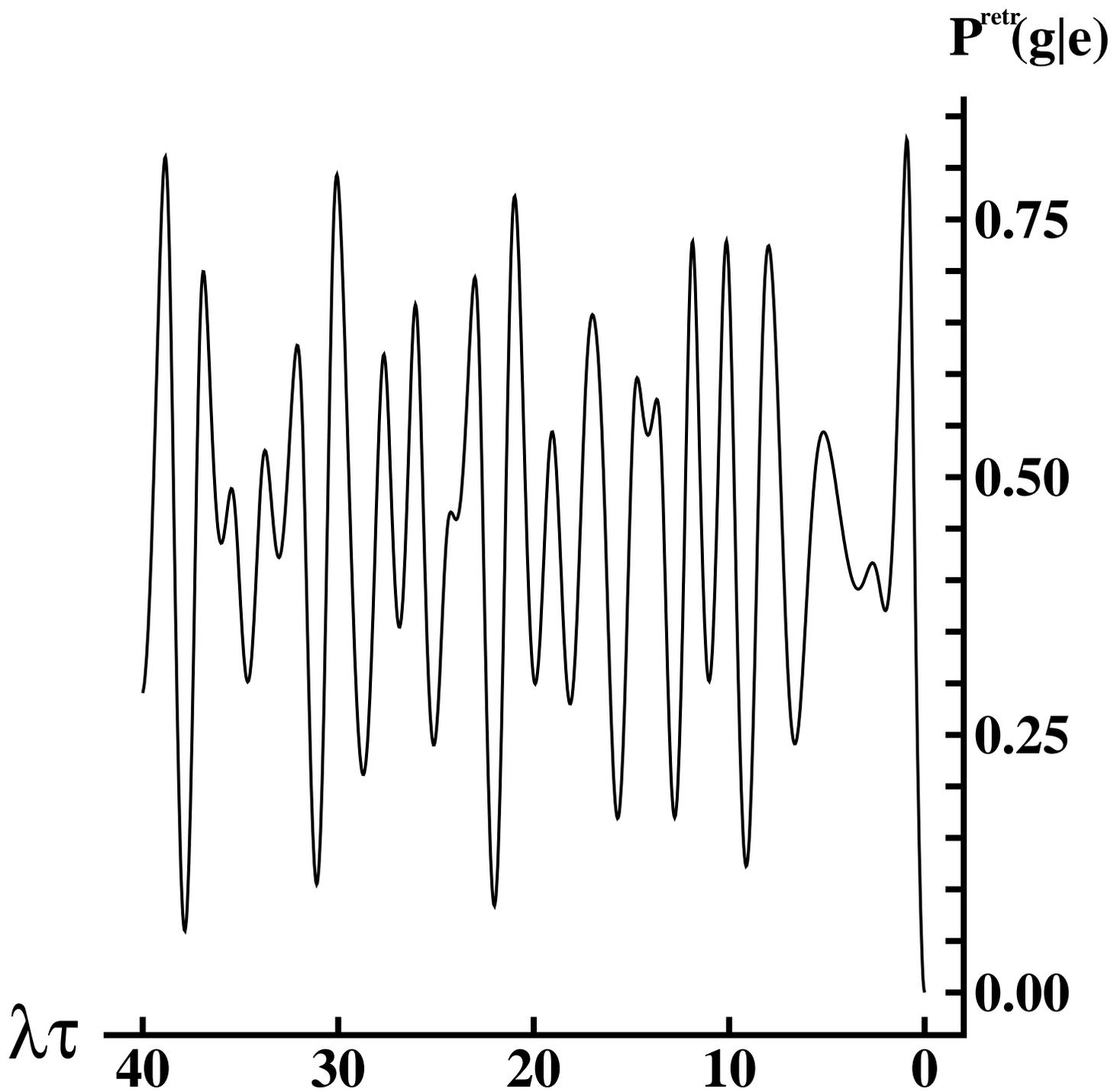
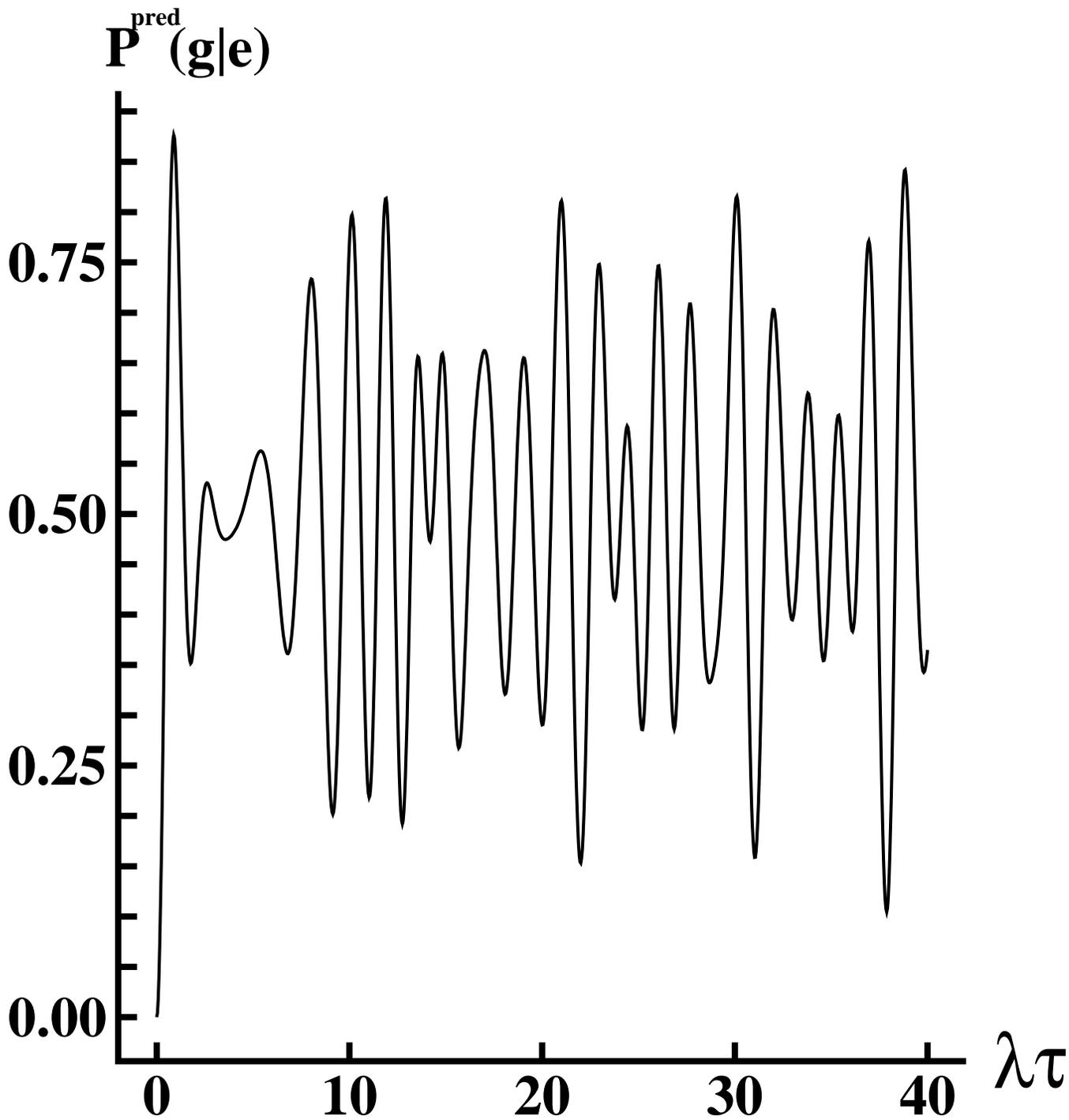
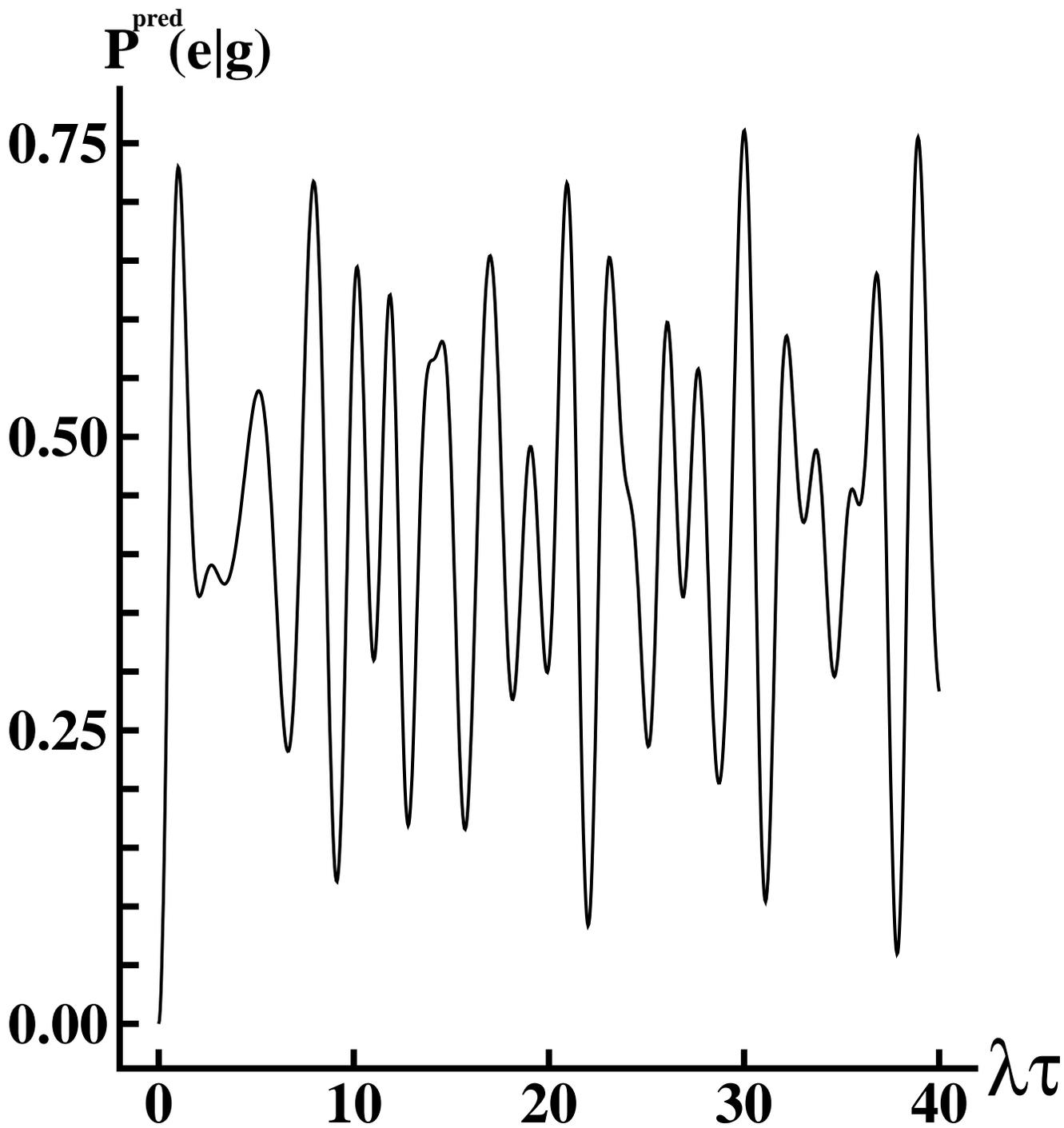
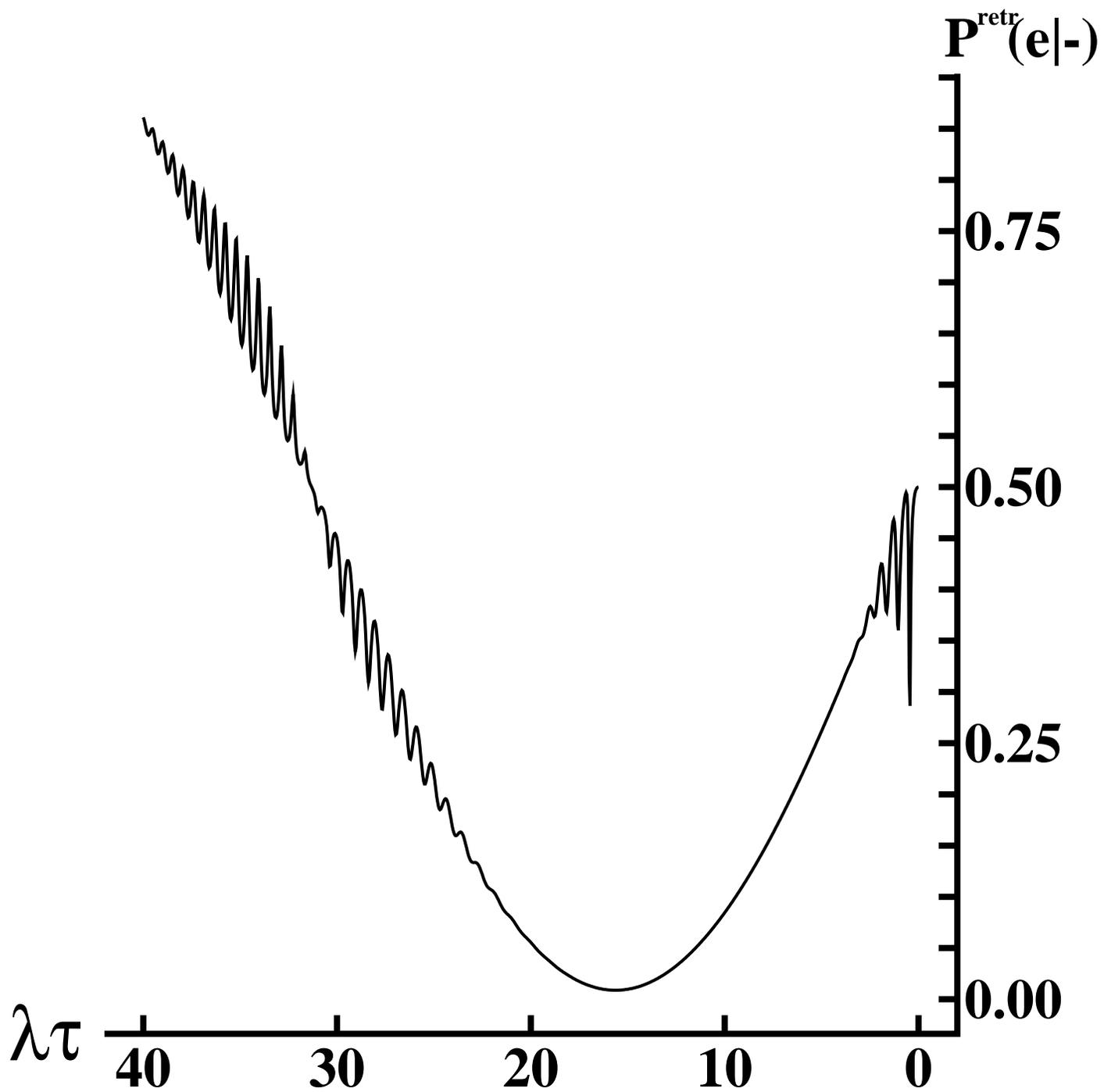
\end{document}